# *Operando* monitoring of single-particle kinetic state-of-charge heterogeneities and cracking in high-rate Li-ion anodes

Alice J. Merryweather[1,2,3], Quentin Jacquet[1], Steffen P. Emge[1], Christoph Schnedermann*[2,3], Akshay Rao*[2,3] & Clare P. Grey*[1,3]

[1] Yusuf Hamied Department of Chemistry, University of Cambridge, Cambridge, UK.

[2] Cavendish Laboratory, University of Cambridge, Cambridge, UK.

[3] The Faraday Institution, Quad One, Harwell Science and Innovation Campus, Didcot, UK

*e-mail:

Christoph Schnedermann: cs2002@cam.ac.uk

Akshay Rao: ar525@cam.ac.uk

Clare P. Grey: cpg27@cam.ac.uk

**Abstract**

Recent years have seen a rapidly escalating demand for battery technologies capable of storing more energy, charging more quickly and having longer usable lifetimes, driven largely by increased electrification of transport and by grid-scale energy storage systems. This has led to the development of many promising new electrode materials for high-rate lithium ion batteries. In order to rationalise and improve upon material performance, it is crucial to understand the fundamental ion-intercalation and degradation mechanisms occurring during realistic battery operation, on the nano- to meso-scale. Here we apply a straightforward laboratory-based *operando* optical scattering microscopy method to study micron-sized rods of the high-rate anode material $Nb_{14}W_3O_{44}$ during cycling at rates of up to 30C. We directly visualise an elongation of the rods, which, by comparison with ensemble X-ray diffraction, allows us to determine the state of charge (SOC) of the individual particle. A continuous change in scattering intensity with SOC is also seen, enabling observation of a non-equilibrium kinetic phase separation within individual particles. Phase field modelling (informed by pulsed-field-gradient nuclear magnetic resonance and electrochemical experiments) is used to verify the kinetic origin of this separation, which arises from a dependence of the Li-ion diffusion coefficient upon SOC. Finally, we witness how such intra-particle SOC heterogeneity can lead to particle cracking; we follow the cycling behaviour of the resultant fragments, and show that they may become electrically disconnected from the electrode. These results demonstrate the power of optical scattering microscopy to track rapid non-equilibrium processes, often occurring over less than 1 minute, which would be inaccessible with established characterisation techniques.

**Introduction**

Lithium-ion batteries that are able to store large amounts of energy in only a few minutes are critical for smart grid systems and to increase the market penetration of electric vehicles. During high-rate charging, electrode materials exhibit non-equilibrium dynamics which include inter- or intra-particle state-of-charge (SOC) heterogeneities,[1–4] polarisation-driven side reactions,[5] electrolyte decomposition,[6] and mechanical degradation involving particle cracking.[7–9] These effects can lower the overall battery performance. Unravelling and understanding these complex non-equilibrium processes in an operating battery is highly challenging.

The development and optimisation of electrode materials suitable for high-rate cycling is stifled by a lack of *operando* characterisation techniques that can monitor Li-ion dynamics at the nanoscale and at relevant charging rates. State-of-the-art electron[8,10,11] and X-ray microscopy techniques[4,12–16] that can access nanometre length scales have led to key insights into the ion transport processes occurring in these systems, *e.g.* by exploring the phase transitions occurring in LiFePO$_4$.[12] However, these single-particle-level imaging techniques are restricted to comparatively slow C-rates of ≤ 2C (where, for $n$C, $n$ is the applied current divided by the theoretical current needed to (dis)charge to a specified capacity in 1 hour).[12] Conversely, ensemble-level characterisation techniques such as electrochemical methods and *operando* X-ray diffraction (XRD) are able to access the relevant time scales, but without being able to spatially resolve these dynamics at the single-particle level.[16–19] Consequently, our understanding of the single-particle ion dynamics at charging rates faster than 2C, while highly desirable, remains largely unexplored.

Under typical cycling conditions, SOC heterogeneity can arise even for 'single-phase' materials which follow a solid-solution mechanism with homogenous Li-concentration at equilibrium. Dynamic non-equilibrium SOC heterogeneities within a solid-solution reaction have previously been evidenced by *ex situ* and ensemble *operando* methods in layered cathode materials during delithiation.[16,19–21] The coexistence of differing SOCs has been attributed to intrinsic kinetic effects, including slow diffusion at high Li-concentrations leading to 'kinetic phase separation' in LiNi$_{0.8}$Co$_{0.15}$Al$_{0.05}$O$_2$ (NCA)[20], and 'autocatalytic' composition-dependant reaction kinetics in LiNi$_{1/3}$Mn$_{1/3}$Co$_{1/3}$O$_2$ (NMC111)[16]. However, without the use of single-particle resolution *operando* techniques, it is difficult to assess how and where these SOC heterogeneities arise during cycling.

Here, we overcome the current lack of understanding by utilising a recently developed optical scattering microscopy technique[22–25] to explore *operando* single-particle ion dynamics at charging rates of up to 30C. We examine Nb$_{14}$W$_3$O$_{44}$ (NWO), which belongs to a family of highly promising fast-charging anode materials,[18,26–28] with record volumetric capacity and charging rates of up to 350 Ah L$^{-1}$ obtained for Nb$_{16}$W$_5$O$_{55}$ at 20C.[18] These high C-rates are substantially too fast to be readily investigated with any other existing *operando* microscopy method, precluding further rationally-guided high-rate optimisations.

By applying *operando* scattering microscopy to rod-shaped NWO particles during cycling at rates from 1C to 30C, pronounced volume changes of the single-particles were directly resolved, which – by comparison with *operando* XRD – allowed us to quantify changes in SOC at the single-particle level throughout cycling. While *operando* XRD performed at a slower rate of C/8 showed only solid solution behaviour, *operando* scattering microscopy revealed phase fronts moving through individual particles, both during the early stages of lithiation from low Li-content and during rapid (> 5C) delithiation from high Li-content, and the velocities of these phase fronts were determined. The phase fronts indicate the emergence of a 'kinetic phase separation', *i.e.* a non-equilibrium spatial variation of SOC within a single particle during cycling. Based on our observations and supporting simulations, we propose that such kinetic phase separation is caused by variation of the Li-ion diffusion coefficient with SOC, and broadly occurs when (dis)charging a 'single-phase' material at sufficiently fast C-rates from a state with comparatively slow Li-ion diffusion towards a state with faster Li-ion diffusion. Finally, we observe a correlation between intra-particle Li-concentration gradients and particle cracking by observing NWO particles cracking in real time, and follow the cycling behaviour of the resultant fragments.

**Material properties and bulk characterisation of NWO**

The crystal structure of $Nb_{14}W_3O_{44}$ is composed of $ReO_3$-like blocks of 4×4 corner-sharing octahedra (Figure 1a). These blocks connect along the *a*- and *b*-axis via crystallographic shear planes, and align along the *c*-axis to form tunnels that enable rapid quasi-one-dimensional Li-ion transport in the *c*-direction.[18,29–31] In this work, two particle morphologies are examined: rod-like particles (typically ~20 μm in size along the *c*-direction and a few microns wide in the *a* and *b*-directions, Figure 1a), and shorter low-aspect-ratio particles (~5 μm in size in all directions, see SI section 1).

Porous self-standing electrodes (170 ± 20 μm thickness) of the rod-shaped particles, sparsely dispersed in a matrix of nanoparticulate carbon and binder, were prepared (see Methods). The electrodes were assembled in a standard coin cell (with a Li metal counter/reference electrode and a carbonate and $LiPF_6$ based electrolyte) and cycled galvanostatically at various C-rates. The resulting sloping voltage-capacity profiles (Figure 1b) are typical of solid-solution (single-phase) intercalation behaviour, and are in good agreement with previous reports.[27] A specific capacity of 150 mA h g$^{-1}$ was achieved at 1C, with capacities of 114 mA h g$^{-1}$ and 83 mA h g$^{-1}$ at faster cycling rates of 5C and 20C, respectively, indicating the high-rate capability and fast lithium-ion diffusion within this material, even in this electrode which has not been optimised for power performance.

To quantify the rate of Li-ion diffusion, pulsed-field-gradient nuclear magnetic resonance (PFG-NMR) experiments at different temperatures (375 K to 465 K) were performed on a lithiated powder with composition $x \approx 0.42$ (in Li$_{17x}$Nb$_{14}$W$_3$O$_{44}$) of low-aspect-ratio particles (see Methods[32], SI sections 2,3,4). The resulting typical Arrhenius-type temperature-dependent diffusion coefficient is plotted in Figure 1c (upper panel). Extrapolating to room temperature results in a self-diffusion coefficient, $D_{self}$, of $(1.8 \pm 0.5) \times 10^{-12}$ m$^2$ s$^{-1}$ with an activation energy of $190 \pm 30$ meV, which is close to values obtained from recent molecular dynamic simulations,[29] and values obtained for Nb$_{16}$W$_5$O$_{55}$.[18] To probe the variation of this room temperature diffusion coefficient as a function of Li-concentration in the material, electrochemical experiments using the galvanostatic intermittent titration technique (GITT) were performed on an electrode of the low-aspect-ratio particles (see Methods). The absolute values of $D_{self}$ determined by GITT depend on the characteristic diffusion length value assumed when analysing the data. Since this value is constant over all SOCs, we used our PFG-NMR results to scale the retrieved GITT diffusion coefficients to appropriate absolute values as a function of Li-concentration (see SI section 5 for further discussion). The resulting composition-dependent self-diffusion coefficient ($D_{self}$, Figure 1c, lower panel) has a maximum value of ~$10^{-11}$ m$^2$ s$^{-1}$, at a composition of $x \approx 0.22$. Towards lower $x$ values, $D_{self}$ drops rapidly by at least 4 orders of magnitude, while a more gradual decrease is observed towards higher $x$, reaching ~$10^{-13}$ m$^2$ s$^{-1}$ at $x \approx 1$. The four orders of magnitude variation of the diffusion coefficient with Li-concentration and the effects on ion transport during rapid cycling are explored further below.

**Optical imaging of NWO single particles**

To understand the high-rate behaviour of individual rod-shaped NWO particles (scanning electron microscopy (SEM) image, Figure 2a), the porous self-standing electrode was assembled into an optically-accessible half-cell (Figure 2b) and galvanostatic discharge-charge cycles were performed at rates of 1C, 5C, 20C and 30C. Unless otherwise specified, each cycle followed a constant voltage hold at 2.8 V vs. Li/Li$^+$, performed at the end of the previous cycle, to ensure a consistent starting Li-concentration (estimated to be $x \approx 0.08$). Throughout these experiments, the electrode surface was examined by *operando* scattering microscopy at a wavelength of 740 nm, while collecting and imaging the light scattered back from the electrode surface (see Methods for a full description of the technique and optical setup).

Figure 2c shows a series of optical images of a 20.4 µm-long NWO particle (as imaged by SEM in Figure 2a), throughout one galvanostatic cycle at 5C (Supplementary Video 1). At all SOCs, the NWO particle appears more brightly scattering than the surrounding carbon matrix, and the observed outline of the particle is replicated from the SEM image. The optical contrast arises primarily from surface reflections, with the ability to detect changes in optical contrast from within the ~320 nm depth of field of our microscope (see SI section 6 for further discussion). Fringe patterns on top and at the edges of the particle arise from optical interference effects, determined by the three-dimensional shape of the particle.

During lithiation, the particle initially became slightly darker before increasing in scattering intensity to become overall more brightly scattering at the end of lithiation compared to the start of the cycle (in this case, the overall intensity increased ~2.1 times). For the majority of the cycle, the intensity changes occurred homogeneously across the whole rod. This trend is reversed upon delithiation, to finish at a scattering intensity close to its initial state. The general intensity behaviour during cycling shown in Figure 2c is typical of all rods and C-rates examined in this study. These observations indicate that the optical scattering intensity of NWO changes non-linearly with the material's SOC and Li-concentration, being primarily sensitive to the electronic structure of the material[33] (further discussion provided in SI section 7).

In addition to changes in scattering intensity, lithiation also caused pronounced volume expansions along the $c$-axis of the rod, visible as a lengthening of the particle, with the reverse occurring during delithiation (Figure 2c). Figure 2d compares the overall cell voltage (top) with the quantitively extracted change in particle length (bottom) at different C-rates, for a similar particle (see SI section 8). This particle's length increased by a total of ~6, 5 and 3% during lithiation at 1C, 5C and 20C after reaching capacities of 180, 106 and 68 mA h g$^{-1}$, respectively. This analysis was extended to several others particles from various electrodes and cycling protocols, for which Figure 2e shows the measured rod elongation as a function of maximum overall cell capacity. The resulting single-particle length expansions are in agreement with Lebail refinement of *operando* XRD of this material at C/8 (grey lines, Figure 2e) and *ex situ* synchrotron XRD (SXRD, grey diamonds, Figure 2e), for which the maximum $c$-parameter expansion was 7% at $x \approx 0.9$ (*i.e.* 160 mA h g$^{-1}$); the XRD results are consistent with previous diffraction studies.[27] It should be noted that the expansions/contractions along the *a*- and *b*-directions[18,27] (see SI section 9) are too small to be optically determined for the sizes of particle in this study.

The good agreement between the *c*-parameter changes and the optically-determined rod elongations demonstrates that changes in Li-concentration at the single-particle level can be monitored by *operando* optical scattering microscopy, enabling heterogeneity in the degree of inter-particle lithium content and reaction heterogeneity as a function of cycling protocol or material morphology to be determined.

**Emergence of kinetic phase separation at low Li-ion concentrations**

For the images presented in Figure 2c, the homogenous intensity changes suggest that the SOC was largely spatially homogenous within the particle for most of the cycle at 5C. However, closer inspection of the cycling behaviour at the very beginning of this cycle (electrochemical data presented in Figure 3a) revealed subtle spatial heterogeneities in intensity. Differential image analysis highlights these features, where each differential image shows the fractional intensity change between a pair of recorded frames separated by a fixed time interval (see SI section 8 for details). Figure 3b shows three such differential images (intensity changes scaled over a 1 s interval) at 3, 6, and 9 s after the start of the galvanostatic lithiation. Initially, once the current was applied, the particle brightened relatively homogenously. At the same time, comparatively dark fronts began moving from both ends of the rod, along the *c*-axis, towards the middle of the particle (Supplementary Video 2). The full evolution of this front movement can be more clearly represented by differential line-cut plots, constructed by selecting a narrow image slice parallel to the direction of the rod length and averaging the differential intensity perpendicularly across the image slice (see SI section 8). Figure 3c shows the resulting differential line-cuts for the first 50 s of this 5C lithiation, clearly highlighting the progression of the fronts. The fronts moved steadily towards the centre of the rod with approximately constant velocities of $600 \pm 70$ nm s$^{-1}$ (from the top end) and $760 \pm 50$ nm s$^{-1}$ (from the bottom end), meeting each other ~15 s after the beginning of the lithiation. Following this, the centre of the particle continued darkening, with this dark central feature spreading to the edges of the rod over time. The intensity changes became homogenous along the length of the particle after about ~40 s, and remained homogenous for the rest of the cycle.

Figure 3d,e,f shows the electrochemical data and the differential images and linecuts of the same particle during a second galvanostatic cycle at 5C, which directly followed the cycle shown in 3a, with no voltage hold in between. Based on the cell electrochemistry, the approximate Li-concentration at the beginning of the cycle was $x \approx 0.14$ (as opposed to ~0.08 for the first cycle, which followed a 2.8 V hold to remove more Li). In contrast to the previous cycle, no clear moving fronts occurred in the particle at the beginning of lithiation and only an initial spatially-homogeneous brightening was observed, albeit lower in intensity (Supplementary Video 2). Whilst the difference in Li-concentration between $x \approx 0.08$ and $x \approx 0.14$ is relatively small, it should be noted that the diffusion coefficient changes dramatically with Li-content at these low values of $x$ (as discussed above, Figure 1c).

Rapid moving fronts during early lithiation following a 2.8 V hold were consistently observed over 17 different particles from 3 different electrodes, at C-rates of 1C, 5C, 20C and 30C, but not in similar cycles without the 2.8 V voltage hold. In no cases were corresponding fronts observed during the final part of delithiation or during the 2.8 V voltage hold.

Spatial heterogeneity in the scattering intensity suggests the presence of SOC heterogeneity within a particle. To explore this possibility, dynamic Li-concentration profiles during galvanostatic cycling were simulated in a 20.4 µm-long NWO rod using a phase field model. The model (see SI section 10) was parameterised using the voltage and diffusion coefficient profiles obtained from GITT and PFG-NMR. Figures 3g and 3h show the simulated differential Li-concentration line-cuts during early lithiation at 5C, representing changes in Li-concentration over time for each location along the rod, starting from uniform compositions of $x = 0.08$ and 0.14, respectively. The simulation starting at $x = 0.08$ reproduces the front-like motion with lithiation fronts moving from the ends of the rod towards the centre at a velocity of 1120 nm s$^{-1}$. After ~20 s, the composition had become uniform within the particle. By contrast, when starting from a higher initial Li-concentration of $x = 0.14$, the degree of SOC heterogeneity within the particle is much less pronounced and more short-lived, reaching a uniform composition after just ~3 s. Equivalent simulations were performed at 1C and 20C, starting from $x = 0.08$, resulting in similar lithiation front behaviour. No thermodynamic phase separation for $0.08 < x < 0.14$ is included in the model, meaning that the SOC heterogeneity is kinetic in origin. This intra-particle SOC heterogeneity can be rationalised by the large change in Li-ion diffusion coefficient in NWO from $x = 0.08$ to $x$

= 0.14 (by ~3 orders of magnitude) which, bearing in mind the wide band-gap (electrically insulating) nature of pristine NWO,[26] is most likely coupled to electronic conductivity limitations of the delithiated material.

Considering the agreement in SOC heterogeneity behaviour between the measured differential intensity line-cuts (Figure 3c,f) and their modelled Li-concentration counterparts (Figure 3g,h), the simulated lithiation front velocities at 1C, 5C and 20C were compared with the experimentally measured velocities at the same cycling rates (Figure 3i). The simulated velocities were 500, 930 and 1900 nm s$^{-1}$, while the mean experimentally observed values were 210 ± 20, 580 ± 30 and 1300 ± 100 nm s$^{-1}$ at 1C, 5C and 20C, respectively. Whilst the simulated front velocities are ~2 times faster than the mean experimentally observed values (see SI section 10 for further discussion), the trend is well reproduced. The optically observed lithiation fronts, accurately reproduced by phase field modelling, validate the existence of kinetic phase separation at high cycling rates in NWO. Additionally, the order-of-magnitude agreement between simulated and observed lithiation front velocities supports our use of the composition-dependent self-diffusion coefficients determined experimentally from GITT and PFG-NMR (Figure 1c) and indicates that they are reasonable estimate of the true diffusion coefficient values at low Li-concentrations in this material.

**Kinetic phase separation leading to particle cracking**

Having established that the sharp drop in ion-diffusivity towards small $x$ can lead to kinetic phase separation during early lithiation, we now turn our attention to exploring the effects of the more gradual decrease in Li-ion-diffusivity towards high values of $x$ (Figure 1c). While imaging a 17.9 μm-long NWO particle (optical image shown in Figure 4a), the electrode was consistently lithiated (to $x \approx 0.68$, based on rod length analysis of the observed particle) at 5C with a 15 min voltage hold at 1.2 V, and then delithiated at C/2, 5C and 20C.

Aside from the observation of the previously-discussed rapid fronts during early lithiation, the scattering intensity changes were homogeneous along the length of the particle during lithiation at 5C, and during delithiation at C/2 (Figure 4b). However, during delithiation at 5C, the differential line-cut analysis displayed in Figure 4c showed the presence of a bright front moving from the bottom of the particle towards the centre (0 – 260 s). After ~280 s of delithiation (to an average of approximately $x \approx 0.56$, based on rod length analysis assuming uniform Li-content), the particle suddenly cracked. We note that the SOC of this particle at the time of cracking corresponds to the point in delithiation with the most rapid changes in the *a*- (and *c*-) lattice parameters (see SI section 9). The new crack remained visible as a faint black line running perpendicularly across the rod (Figure 4d).

Figure 4e shows differential images of the initial bright front, which moved at an average speed of $44.2 \pm 0.3$ nm s$^{-1}$, becoming gradually more diffuse over the first ~260 s of delithiation. Differential images of the cracking process (Figure 4f, Supplementary Video 3) show that the dark crack initially started at the left-hand side of the particle, from which it propagated perpendicularly across the rod at a velocity of $187 \pm 4$ nm s$^{-1}$. During and after the formation of the crack, additional front-like features began to spread out from the crack along the length of the two fragments. This indicates that electrolyte was able to penetrate the crack and the newly cleaved surfaces acted as active surfaces for the remainder of the delithiation.

The electrode was then relithiated to $x \approx 0.68$ and delithiated at 20C. Sharp bright delithiation fronts propagating from the bottom of the rod (at $85 \pm 1$ nm s$^{-1}$) and from both sides of the crack (at $143 \pm 2$ and $128 \pm 1$ nm s$^{-1}$) were observed (Figure 4g,h, Supplementary Video 4), clearly demonstrating that the cracked surfaces were active surfaces for delithiation. We note that delithiation fronts do not propagate from the top of this rod at any of the applied C-rates, suggesting that (in contrast to the particle presented in Figure 3) the top end does not act as an active surface in this particle. At 75 s, continued volume contraction caused the fractured particle to further separate into two distinct pieces (shown in Figure 4i, following completion of the delithiation). Further cycling resulted in additional cracking of this particle (Figure 4j).

The heterogeneous scattering intensity during rapid delithiation from $x \approx 0.68$ is again indicative of underlying heterogeneity in Li-concentration, with the observed rate-dependence confirming a kinetic origin. In this case, at 5C, intra-particle heterogeneity caused cracking, likely due to severe strain arising from spatial heterogeneity in the crystal lattice parameters. Similar cracking was observed in real time for four different NWO particles, always occurring during delithiation and not during lithiation.

For NWO, cracking allows the formation of new active surface. However, cracking can also lead to significant particle disconnection. Figure 5a shows optical scattering microscopy images of selected cracked NWO particles in a delithiated electrode following cycling. In each image, certain particle fragments are distinctly more brightly scattering compared to other fragments and neighbouring particles, suggesting that they had become disconnected from the electrode and contain Li that has remain trapped (Figure 5b). This highlights the need to ensure good electronic conduction pathways within an electrode, and future work will include more detailed investigations into cracking-induced particle disconnection, including quantifying its impact on overall cell capacity fade.

**Discussion**

Taken together, our optical studies revealed two distinct regimes of kinetic phase separation in NWO: during initial lithiation starting from low Li-concentration (Figure 3), and during rapid (>5C) delithiation starting from high Li-concentration (Figure 4). In both cases, experiments and simulations suggest that this phase separation is only observed when (dis)charging the material from a state with comparatively slow diffusion towards a state with faster diffusion. In addition, the rate of ion (de)insertion (*i.e.* the applied current) must be faster than the ion diffusion in the initial low-mobility phase. Only when both of these conditions are met, does the new higher-mobility phase continue to accumulate near the active surfaces faster than ions in the lower-mobility bulk can diffuse towards a uniform equilibrium composition, resulting in non-equilibrium Li-concentration heterogeneity within individual active particles and a 'kinetic phase separation'.

Intra-particle kinetic phase separation is likely relevant to many other nominally 'single-phase' materials besides NWO, as suggested by recent ensemble and *ex situ* studies of NCA and NMC.[16,19,20] However, to the best of our knowledge, our work represents the first time that kinetic phase separation has been directly visualised under *operando* conditions and at the single-particle level, confirming the existence of SOC-heterogeneity within individual particles of nominally 'single-phase' materials.

When coexisting phases have substantially different lattice parameters, phase separation within particles leads to internal strain which may in turn cause particle fracture, as observed here during delithiation of NWO. We note that particle cracking has not been observed during the kinetic two-phase behaviour at early lithiation, where the coexisting phases have more similar Li-concentrations and lattice parameters. Particle fracture is widely accepted to be one significant cause of electrode capacity fade across many electrode chemistries due to electrical disconnection of particles,[7–9] highlighting the critical importance of understanding cracking and its effect on particle connectivity. Future work will examine the origin of intra-particle strain in more detail, as well as investigating the effects of defects and grain boundaries – including intergrowths of alternative crystal phases – upon particle cracking.

**Conclusions**

Using *operando* optical scattering microscopy to examine individual rod-like particles of NWO, we observed an overall increase in scattering intensity and a rod-length expansion during lithiation, with the reverse occurring during delithiation. By correlating the observed length expansion/contraction with *c*-lattice parameter changes measured by *operando* and *ex situ* XRD, we have shown that changes in SOC can be optically monitored and quantified in individual particles.

Ensemble PFG-NMR experiments determined a Li-ion diffusion coefficient on the order of $10^{-12}$ m² s$^{-1}$ for $x \approx$ 0.42, while GITT experiments revealed that this diffusion coefficient varies significantly with SOC, dropping sharply at low $x$ and decreasing more gradually at high $x$. Using *operando* optical scattering microscopy, we identified kinetic phase fronts moving through individual particles during the early stages of lithiation from low Li-content and during delithiation from high Li-content at rapid C-rates. Fickian diffusion modelling using the experimentally-determined composition-dependent diffusion coefficient reproduced the observed phase front behaviour, including a good agreement with their velocities at a range of C-rates, and confirming the kinetic origin of the two-phase behaviour.

Finally, we observed NWO particles cracking in real time, following the build-up of intra-particle SOC heterogeneity during rapid delithiation from high Li-content. Some of the resultant fragments were observed to continue to (de)lithiate during subsequent cycles akin to the non-broken particles, while others became electrically disconnected from the electrode.

These results highlight the ability of our imaging method to effectively capture rapid dynamic processes (often occurring over less than 1 min) and to provide real-time insights into nanoscale ion transport and particle cracking behaviour. Such studies are not feasible with conventional ensemble characterisation methods, which measure only the average behaviour across an entire electrode. *Operando* imaging methods with single-particle resolution and rapid (< 1 s) acquisition times are essential in observing non-equilibrium processes on the relevant lengthscales and timescales. Due to its broad applicability to a wide range of battery materials and its relatively straightforward implementation, our optical scattering microscopy method promises to be highly valuable in advancing future understanding of fundamental ion transport and electrode degradation mechanisms.

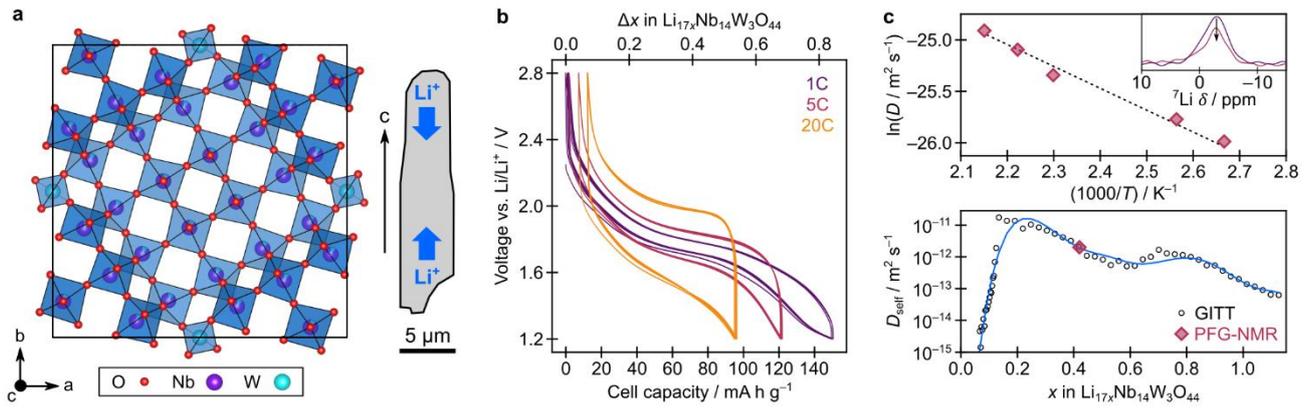

**Figure 1: Structure, cycling performance and Li-diffusion coefficients of NWO.**
**a** Left: Crystal structure of the $Nb_{14}W_3O_{44}$ unit cell. Right: Schematic of a rod-shaped particle of NWO. The long direction of the rod is the crystallographic *c*-axis, the direction along which Li-ion transport occurs. **b** Specific capacity plots for a self-standing NWO electrode in a coin cell, cycled galvanostatically from 1.2 – 2.8 V at rates of 1C (purple), 5C (red) and 20C (orange), for five cycles at each rate. **c** Top: Arrenhius plot showing the variation of the Li self-diffusion coefficent ($D_{self}$) for $x \approx 0.42$ (in $Li_{17x}Nb_{14}W_3O_{44}$) as a function of temperature, obtained from $^7$Li PFG-NMR with short diffusion time ($\Delta$ = 10 ms) to avoid a constrained diffusion regime. Extrapolation to room temperature yields $D_{self}(25°C) = (1.8 \pm 0.5) \times 10^{-12}$ m$^2$ s$^{-1}$, with an activation energy of $E_a = 190 \pm 30$ meV (Inset: Typical $^7$Li NMR signals, with a black arrow indicating a reduction in peak intensity with increasing magnetic field gradient) Bottom: Black cicles show the variation of $D_{self}$ as a function of Li-concentration, obtained from GITT, with the blue line representing a polynomial fit to the data points. The absolute values of these GITT results are scaled to match the room temperature value of $D_{self}$ for $x \approx 0.42$, obtained from PFG-NMR, which is indicated by a red diamond.

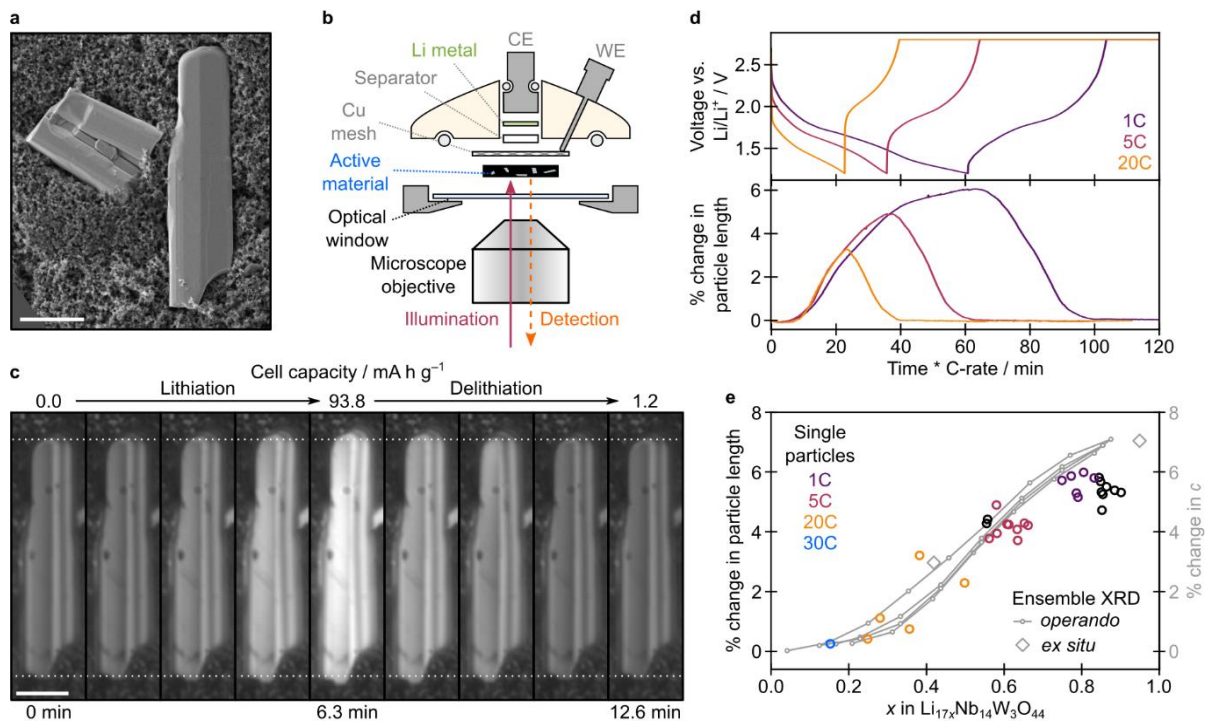

**Figure 2: Optical response and volume expansion of NWO particles during cycling.**
**a** SEM image of two typical rod-shaped NWO particles in an electrode. Scale bar is 5 μm. **b** Geometry of the optical microscopy half-cell. (WE = working electrode, CE = counter electrode). The counter electrode was lithium metal, the separator was glass fibre, and the cell stack was wet with standard carbonate liquid electrolyte (LP30). **c** Optical scattering images of the longer active particle at evenly spaced time-points throughout a galvanostatic cycle at 5C (Supplementary Video 1). Intensity values are normalised to a linear grey-scale between 0 (black) and 1 (white). The dotted white lines are a guide to the eye indicting the initial length of the particle. Scale bar is 5 μm. **d** Top: Cell voltage during galvanostatic cycles at 1C (purple), 5C (red) and 20C (orange), each followed by a constant voltage hold at 2.8 V. Bottom: Relative length expansion of an individual NWO particle, determined from *operando* optical imaging, during the three cycles shown in the top panel. **e** Each hollow circle indicates the overall length expansion of an individual NWO particle determined from *operando* optical imaging, as a function of the maximum cell (charge) capacity achieved. Cycles at 1C, 5C, 20C and 30C are shown in purple, red, orange and blue respectively, while black indicates non-standard cycling protocols (*e.g.* including additional constant voltage holds). The crystallographic *c*-lattice expansions determined from ensemble *operando* XRD (C/8, 1st and 2nd cycles, grey lines) and *ex situ* SXRD (grey diamonds) are also shown.

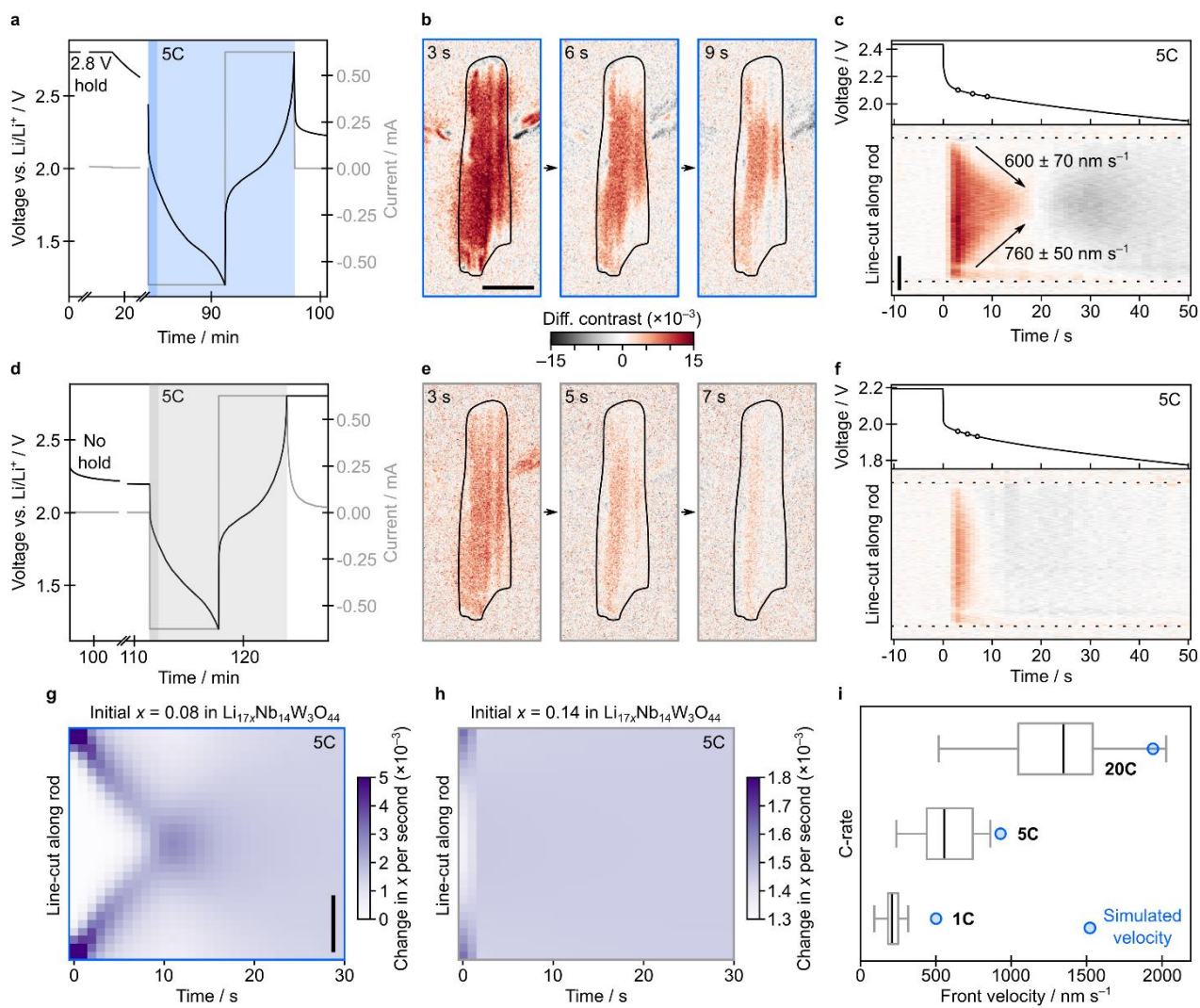

**Figure 3: Rapid phase fronts at the beginning of lithiation from low Li-concentration.**
**a** Cell voltage (black line) and current (grey line) during a galvanostatic cycle at 5C (light blue rectangle), following a constant voltage hold at 2.8 V. **b** Differential images of the active NWO particle at 3, 6 and 9 s after the start of lithiation (Supplementary Video 2). The differential contrast represents the fractional intensity changes per second, at each pixel. Panels b,c,e,f share the same colour-scale. **c** Top: Cell voltage over the first 50 s of lithiation. Time-points corresponding to the images shown in b are indicated with small open circles. Bottom: Corresponding evolution of a differential line-cut along the length of the rod ($y$-axis) with time ($x$-axis). Dotted black lines indicate the positions of the particle edges. Black arrows are a guide to the eye, highlighting the phase front motion. **d** Cell voltage (black line) and current (grey line) during a galvanostatic cycle at 5C (light grey rectangle), following immediately from the cycle shown in a. **e** Differential images of the active NWO particle at 3, 5 and 7 s after the start of lithiation (Supplementary Video 2). **f** Top: Cell voltage over the first 50 s of lithiation. Time-points corresponding to the images shown in e are indicated with small open circles. Bottom: Corresponding evolution of a differential line-cut along the length of the rod. **g,h** Simulated rate of change of Li-concentration along a 1D NWO rod ($y$-axis) with time ($x$-axis), for a 5C lithiation from initial states of $x = 0.08$ and 0.14, respectively. **i** Box and whisker plots show the spread of phase front velocities measured for different rod-shaped NWO particles at applied C-rates of 1C, 5C and 20C, based on 16, 41 and 16 different velocity values respectively. Blue circles represent the simulated front velocities at the same C-rates. All scale bars are 5 μm.

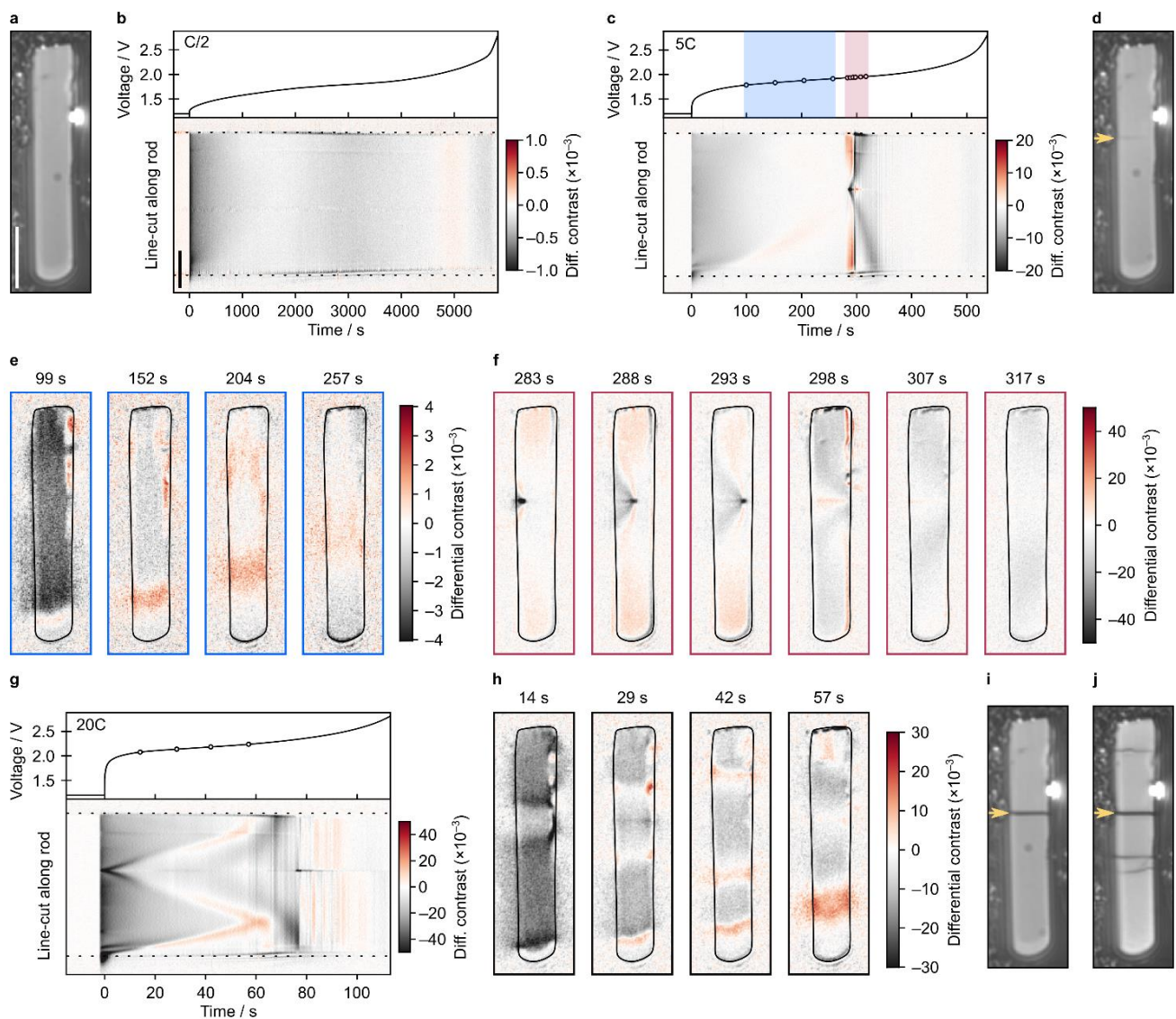

**Figure 4: Kinetic phase separation during delithiation, leading to particle cracking.**
**a** Optical scattering image of an active rod-shaped NWO particle. Intensity values are normalised to a linear grey-scale between 0 (black) and 1 (white). **b** Top: Cell voltage during delithiation at C/2, from an initial state of $x \approx 0.68$ (for the observed particle). Bottom: Corresponding evolution of a differential line-cut along the length of the rod ($y$-axis) with time ($x$-axis). The differential contrast represents the fractional intensity changes per second, at each pixel. Dotted black lines indicate the initial positions of the particle edges. **c** Top: Cell voltage during delithiation at 5C, from an initial state of $x \approx 0.68$. Time-points corresponding to the images shown in e and f are indicated with small open circles. Bottom: Corresponding evolution of a differential line-cut along the length of the rod. **d** Optical scattering image of the same active particle after the 5C delithiation. **e,f** Differential images of the active NWO particle during the 5C delithiation (Supplementary Video 3). In addition to moving fronts, changes in scattering intensity can be observed at the top-right side of the particle (99, 152, 204 s) mostly likely due to uneven surface morphology providing additional active surface. **g** Top: Cell voltage during delithiation at 20C, from an initial state of $x \approx 0.68$. Time-points corresponding to the images shown in h are indicated with small open circles. Bottom: Corresponding evolution of a differential line-cut along the length of the rod. **h** Differential images of the active NWO particle during the 20C delithiation (Supplementary Video 4). **i** Optical scattering image of the same active particle after the 20C delithiation. **j** Optical scattering image of the same active particle following an additional 9 cycles covering a range of cycling protocols (including further rapid delithiation from high Li-concentrations).

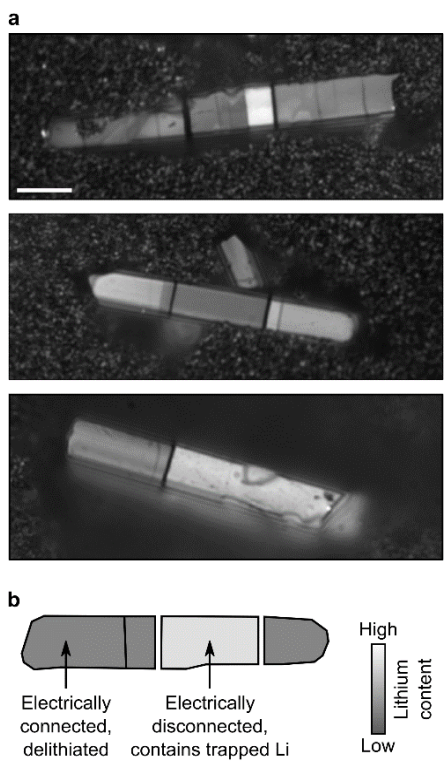

**Figure 5: Electrical disconnection of cracked NWO particles.**
**a** Optical scattering microscopy images of selected NWO particles in a delithiated electrode, following a total of 20 complete lithiation-delithiation cycles, including rapid (up to 20C) delithiation from high Li-concentrations. Dark lines/gaps running across these particles indicate that they are broken into multiple fragments, some of which appear substantially more brightly scattering. The mean scattering intensities of the brighter segments are ~1.7, ~1.4 and ~1.4 times higher than for the darker segments in the top, middle and bottom images, respectively. All three images are displayed with the same scale (5 μm scale bar) and intensity range (normalised to a linear grey-scale between black and white). **b** Schematic of a similar NWO particle, illustrating that the more brightly scattering fragments likely have a higher Li-content and became electrically disconnected from the rest of the electrode.

**Methods**

1. **Sample preparation:**

*Synthesis of $Nb_{14}W_3O_{44}$*

Niobic acid (HY340-CBMM) and $WO_3$ (Sigma-Aldrich) were ball milled for 1 hour in stoichiometric amounts in a SPEX, using a 45 mL zirconia jar. The typical loading was 1 g of precursor with three 10 mm zirconia balls. The ground precursors were heated at 700°C for 6 hours, directly followed by 1200°C for 12 hours (using a 10°C/min heating ramp). The cooling was not controlled. At this stage, low-aspect-ratio NWO particles were obtained. The longer rod-like NWO particles were obtained by mixing equal masses of the low-aspect-ratio sample with additional ground precursor, and repeating the heat treatment (700°C for 6 hours and 1200°C for 12 hours).

*Electrochemical ex situ sample preparation*

*Ex situ* samples for PFG and SXRD were prepared by pressing around 250 mg of low-aspect-ratio $Nb_{14}W_3O_{44}$ particles into a pellet and sintering at 500°C for 2 hours. The pellet was assembled in a Swagelok-type electrochemical cell with a glass fibre separator (Whatman, GF/B glass microfiber filter) soaked in LP30 liquid electrolyte (Sigma-Alrich, 1 M $LiPF_6$ in a 50:50 mixture of ethylene carbonate and dimethyl carbonate). The counter electrode was lithium metal. The pellet was then lithiated by a series of constant voltage steps (see SI Figure 2). After reaching the desired lithiation state, the pellet was removed from the cell and rinsed thoroughly in DMC to remove any residues of $LiPF_6$.

*Electrode preparation for operando optical scattering microscopy and XRD*

Self-standing electrode films were prepared from NWO, Super P™ carbon and polytetrafluoroethylene suspension (PTFE, 60% in water), combined by grinding together with ethanol in a pestle and mortar. For characterisation by optical scattering microscopy, the dry mass ratios of long rod-like NWO particles, carbon and PTFE were 20%, 46% and 34%, respectively. For characterisation by XRD, the dry mass ratios of low-aspect-ratio NWO, carbon and PTFE were 80%, 10% and 10%, respectively. Upon drying to a workable consistency, the mixtures were rolled flat into porous self-standing films with a thickness of ~170 ± 20 μm. The films were fully dried (vacuum oven, 100°C, overnight) before cutting electrode disks (5 mm diameter for optical scattering microscopy, 20 mm diameter for XRD). The electrodes were stored in a dry argon atmosphere prior to use.

2. **X-ray diffraction:**

*Operando* XRD was performed using an electrochemical cell with Be windows.[34] Patterns were obtained at 10 min intervals during galvanostatic (de)lithiation at C/8, using a Bragg–Brentano geometry (Empyrean, Panalytical) at ambient temperature with a Cu-Kα source. (Throughout this work, for *n*C, *n* is the applied current divided by the theoretical current needed to (dis)charge to a specified capacity of 178 mA h g$^{-1}$ in 1 hour). The *ex situ* samples were sealed in 0.5-mm glass capillaries to prevent exposure to air, and synchrotron XRD was performed at 11-BM (APS) in transmission geometry. All Lebail refinements were performed using Fullprof.[35]

3. **PFG-NMR:**

PFG-NMR experiments were carried out using a Bruker 300 MHz (7.1 T) magnet with an Avance III console and a Diff-50 probe head equipped with an extended variable-temperature 5-mm single-tuned $^7$Li saddle coil insert. Spectra were recorded with a stimulated-echo PFG sequence to minimize $T_2$ losses. In order to avoid rapid $T_2$ relaxation at room temperature, PFG-NMR was performed at temperatures ranging from 375 K to 465 K, which is within the thermal stability window of lithiated NWO (confirmed by XRD, see SI section 3). The 'opt' shaped gradient pulse maximum strength and duration, $\delta$, were 2300 G cm$^{-1}$ and 2 ms, respectively. Long delays (2 ms) were used between PFG pulses and measurements to minimize eddy currents, and as a result no dephasing is observed in the spectra (SI Figure 4f) with the increase of the magnetic field gradient. Diffusion time, $\Delta$, was varied between 7 and 100 ms. *Ex situ* samples of the isotopically-shaped NWO particles were prepared as detailed above. The low-aspect-ratio particle morphology was chosen to reduce experimental complications which could arise from having a strong preferred orientation of high aspect-ratio (rod-like) particles in the sample. The sample was packed into a 5-mm glass NMR tube, sealed with an air-tight J Young NMR valve. No changes were observed when comparing the $^7$Li static spectra before and after the PFG sequences. The recycle delay between pulse sequences was set to 1 s. The gradient strength was tested and calibrated on a saturated LiCl solution.

As indicated by recent DFT calculations and the structures of these materials, Li diffusion in NWO was considered to be 1D (along the tunnels in the crystal $c$-direction, see Figure 1a)[29]. Li-ion self-diffusion coefficients ($D_{self} = \langle z^2(\Delta) \rangle/\Delta$) at various temperatures were obtained from the PFG decay over short diffusion times ($\Delta$ = 10 ms), by fitting the decay using a 1D diffusion model. Theoretically, the expression for the PFG-NMR spin echo attenuation in the case of 1D diffusion is given by equation 1,

$$\Psi(\delta,\Delta,g) = \frac{1}{2}\int_0^\pi \exp\left(-\frac{1}{2}(\gamma\delta g)^2 \langle z^2(\Delta)\rangle \cos^2\theta\right)\sin\theta\, d\theta, \quad (1)$$

where $\Psi$, $\delta$, $\Delta$, $g$, $\gamma$, $\theta$ and $\langle z^2(\Delta) \rangle$ are the spin echo attenuation, the pulse duration, the diffusion time, the gradient strength, the gyromagnetic ratio, the angle between the diffusion channel and the magnetic field gradient, and the mean square displacement along the direction of the magnetic field gradient, respectively.[32] Equation 1 assumes that the directions of the 1D diffusion tunnels follow a statistical isotropic powder-average orientation in space, in other words, that there is no preferred orientation of the particles in the sample. For fitting the results, the integral in equation 1 was approximated numerically using a sum of 89 angles selected using the ZCW scheme[36–38] performed using a MATLAB script.

4. **GITT:**

Electrochemical experiments using the galvanostatic intermittent titration technique (GITT) were performed using the low-aspect-ratio NWO in a thin and porous electrode (2.3 mg cm$^{-2}$ loading; 80% NWO, 10% Super P™ carbon, 10% polyvinylidene difluoride) coated on Cu foil. The GITT experiment was performed by applying a series of C/10 constant-current pulses, each lasting 30 minutes, with a 5-hour relaxation period between each pulse. Data points were recorded at 1 mV intervals (SI Figure 5a). At short times after switching on the current pulse, the Fick equation predicts the cell potential to vary linearly with the square root of time (SI Figure 5b-e) with the slope giving access to the diffusion coefficient according equation 2:

$$\frac{D_{\text{chem}}}{L^2} = \frac{4}{\pi}\left(\frac{I}{Z_a F}\right)^2 \left(\frac{dE(x)}{dx}\right)^2 \left(\frac{d\sqrt{t}}{dE(x)}\right)^2, \quad (2)$$

where $D_{\text{chem}}$, $L$, $I$, $Z_a$, $F$, $E$, and $x$ are the chemical diffusion coefficient, the diffusion length, the applied current, the charge number of the electroactive species ($Z_a = 1$ for $Li^+$), the Faraday constant, the cell voltage and the amount of lithium in $Li_{17x}Nb_{14}W_3O_{44}$, respectively. A good agreement of the variation of the reciprocal time constant $D_{\text{chem}}/L^2$ with the voltage was obtained for two cells during lithiation and delithiation. Note that analysis at long times after switching on the current pulse ($t \gg L^2/D_{\text{chem}}$) was also carried out and was in agreement with the results for short times.

### 5. Optical scattering microscopy:

*Optical setup*

Optical scattering microscopy was carried out by adapting our previously described microscope setup.[22] Briefly, a home-built inverted interferometric scattering microscope equipped with an oil immersion objective (100×, UPLSAPO100XO, Olympus) and polarisation optics in the detection path imaged the reflected and scattered light from the sample onto a CMOS detector (FLIR, Grashopper3, GS3-U3-23S6M-C) with an overall magnification of 166.7×. In this work, the sample was illuminated at 740 nm by a high power LED source (Thorlabs SOLIS-740C), equipped with a ground glass diffuser to minimise speckle contributions and homogenise the illumination. The field of view was controlled by a field aperture and set to achieve a circular illumination profile with a diameter of 35 µm.

The sample cell was mounted on an XYZ nano-positioner stack (Attocube, ECSx3030/AL/RT/NUM) with an overall travel range of 25 mm in all dimensions. The sample focus position was maintained to within 4 nm via an active external focus stabilisation based on a calibrated line-reflection profile of a 980 nm reference laser, as described previously.

*Data acquisition*

Optical scattering studies were carried out to examine 17 different rod-shaped NWO particles, in 3 different electrodes, using a range of cycling protocols and data acquisition parameters. For the experiments presented in Figures 2c, and 3b,c,e,f, images were acquired with a camera exposure time of 8 ms, at a frame rate of 2 Hz. Each recorded image was spatially binned (2×2 pixels, giving an effective pixel size of 69.4 nm $px^{-1}$) and sets of 2 recorded images were temporally binned together to yield an effective frame rate of 1 Hz. For the data presented in Figure 4, the exposure time was 7 ms, the frame rate was 10 Hz, and each image was spatially binned (2×2 pixels). For the 5C and 20C delithiation experiments, sets of 5 images were temporally binned (2 Hz effective frame rate), while sets of 15 were used for the C/2 delithiation (0.67 Hz effective frame rate).

Electrochemical control was achieved using a Gamry potentiostat (Interface 1010). Image acquisition and synchronisation of instruments was performed using in-house developed LabVIEW routines.

*Image analysis*

See SI section 8 for details of the image analysis procedures.

## 6. Phase field modelling

A simplified phase field model was used to simulate the Li-concentration profile in typical NWO particles during battery cycling (see SI section 10 for further description). The inputs to the model, *i.e.* the Li-ion self-diffusion coefficient and the chemical potential, were parameterised from GITT and PFG-NMR experiments. For the simulations presented in this work, the evolution of the Li-concentration profile with time was calculated for a 20.4 µm-long NWO rod (with a cross section of 3.5 × 3.5 µm) during (de)lithiation at C-rates ranging from 1C to 20C.

**Acknowledgements**

This work was supported by the Faraday Institution (FIRG012 Characterisation). A.J.M. acknowledges support from the EPSRC Cambridge NanoDTC, EP/L015978/1. C.S. acknowledges financial support by the Royal Commission of the Exhibition of 1851. We acknowledge financial support from the EPSRC and the Winton Program for the Physics of Sustainability. This project has received funding from the European Research Council (ERC) under the European Union's Horizon 2020 research and innovation program (Grant Agreement No. 758826). C.P.G., S.P.E and A.J.M. were supported by an ERC Advanced Investigator Grant for Prof. Clare Grey (EC H2020 835073). Use of the Advanced Photon Source at Argonne National Laboratory was supported by the U. S. Department of Energy, Office of Science, Office of Basic Energy Sciences, under Contract No. DE-AC02-06CH11357. We thank S. Nagendran and J. Thuillier for their help with synthesizing the materials and with the PFG-NMR measurements, P. Magusin for advice regarding the PFG-NMR measurements, and B. Mockus for help with the code development. C.P.G. is a cofounder of the company Nyobolt (https://nyobolt.com) that seeks to develop fast charging batteries for a variety of applications.

**Author contributions**

C.P.G and A.R. conceived the idea, and supervised the project. A.J.M., Q.J. and C.S. planned all experiments. Q.J. and A.J.M. prepared samples. A.J.M. and C.S. constructed the optical setup and carried out the optical measurements. Q.J. performed the XRD and GITT measurements and the modelling. Q.J and S.P.E carried out the PFG experiments. All authors discussed the results and contributed to writing the manuscript.

**Competing interests**

The authors declare no competing interests.

**Additional information**

Supplementary Information is available for this paper.

Correspondence and requests for materials should be addressed to:

Christoph Schnedermann: cs2002@cam.ac.uk

Akshay Rao: ar525@cam.ac.uk

Clare P. Grey: cpg27@cam.ac.uk

**Data availability**

The data underlying all figures in the main text and Supplementary Information are publicly available from the University of Cambridge repository.

**Code Availability**

All code used in this work is available from the corresponding author upon reasonable request.